\documentclass{elsart}

\usepackage{graphicx}
\usepackage{amsmath}
\usepackage{amssymb}

\begin{document}

\journal{Nucl. Inst. and Meth. A}


\begin{frontmatter}

\title{Mass production test of Hamamatsu MPPC for T2K neutrino oscillation experiment}
\author[KYOTO]{M.~Yokoyama\corauthref{cor}},
\author[KYOTO]{T.~Nakaya},
\author[KYOTO]{S.~Gomi},
\author[KYOTO]{A.~Minamino},
\author[KYOTO]{N.~Nagai},
\author[KYOTO]{K.~Nitta},
\author[KYOTO]{D.~Orme},
\author[KYOTO]{M.~Otani},
\author[KEK]{T.~Murakami},
\author[KEK]{T.~Nakadaira},
\author[KEK]{M.~Tanaka}

\address[KYOTO]{Department of Physics, Kyoto University, Kyoto, 606-8502 Japan}
\address[KEK]{IPNS, High Energy Accelerator Research Organization (KEK), Tsukuba, Ibaraki 305-0801, Japan}

\corauth[cor]{Corresponding author.\\ \texttt{masashi@scphys.kyoto-u.ac.jp}}

\begin{abstract}
In the T2K near neutrino detectors, about 60 000 Hamamatsu Multi-Pixel Photon Counters (MPPCs) will be used. 
The mass production of MPPC has started in February 2008.
In order to perform quality assurance and to characterize each device, we have developed an MPPC test system. 
For each MPPC, gain, breakdown voltage, noise rate, photo detection efficiency, and cross-talk and after-pulse rate are measured as functions of the bias voltage and temperature. 
The design of the test system and the measurement procedure are described.
\end{abstract}

\end{frontmatter}

\section{Introduction}
The T2K (Tokai-to-Kamioka)~\cite{T2K} is a long baseline neutrino oscillation experiment in Japan.
In the near detector system~\cite{T2K-ND280}, we plan to use about 60~000 Hamamatsu Multi-Pixel Photon Counters (MPPCs)~\cite{MPPC, MPPC2}.

\begin{table}[htbp]
\begin{center}
\caption{Specifications of T2K-MPPC (S10362-13-050C).}
\begin{tabular}{ccc} \hline \hline
\multicolumn{2}{c}{Item} & Spec. \\ \hline
\multicolumn{2}{c}{Active area} & 1.3$\times$1.3~mm$^2$ \\
\multicolumn{2}{c}{Pixel size} & 50$\times$50~$\mu$m$^2$ \\
\multicolumn{2}{c}{Number of pixels} & 667 \\
\multicolumn{2}{c}{Operation voltage} & 70~V (typ.) \\
\multicolumn{2}{c}{PDE @ 550~nm} & $>$15\% \\
Dark count &($>$0.5~pe) & $<$1.35~Mcps\\
$[$ @ 25$^\circ$C $]$  & ($>$1.2~pe) & $<$0.135~Mcps \\
\hline \hline
\end{tabular}
\label{tab:spec}
\end{center}
\end{table}

\begin{figure}[tbp]
\begin{center}
\includegraphics[width=0.4\textwidth]{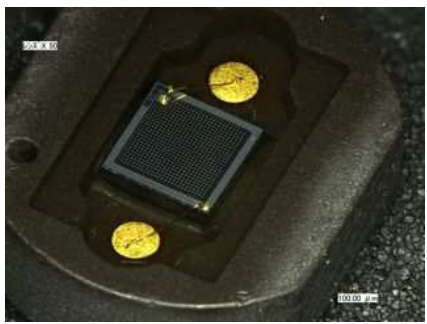}%
\caption{Picture of T2K-MPPC.}
\label{fig:MPPC}
\end{center}
\end{figure}%

We have worked with Hamamatsu to develop the MPPC for T2K in the past a few years.
The major specifications of T2K-MPPC are summarized in Tab.~\ref{tab:spec}.
A picture of T2K-MPPC is shown in Fig.~\ref{fig:MPPC}.
The mass production started in February 2008.
In order to perform quality assurance and to characterize each device, we have developed an MPPC test system. 
By August 2008, 15~000 MPPCs will be tested with our system at Kyoto University, while the remainder will be tested at other testing facilities abroad.

In this paper, the design of the test system, testing procedure and performance of T2K-MPPC are presented.

\section{Test system}

\begin{figure}[htbp]
\begin{center}
\includegraphics[width=0.48\textwidth]{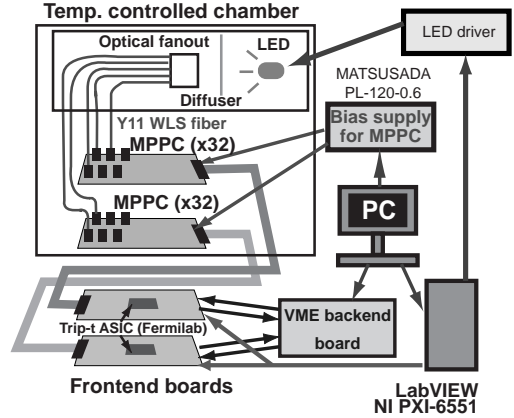}%
\caption{Schematic diagram of the MPPC test system.}
\label{fig:setup}
\end{center}
\end{figure}%

\begin{figure}[htbp]
\begin{center}
\includegraphics[width=0.45\textwidth]{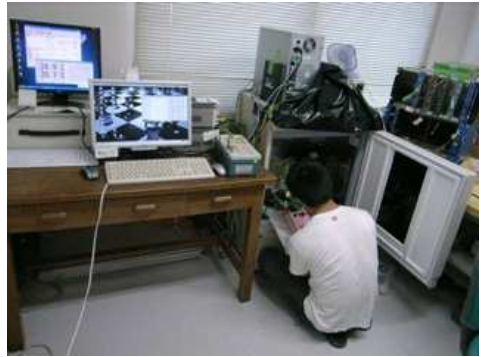}%
\caption{Picture of the MPPC test system.}
\label{fig:setup-pict}
\end{center}
\end{figure}%

In order to characterize a large number of MPPCs, we have developed a system that can simultaneously measure 64 MPPCs.
Figure~\ref{fig:setup} shows the schematics of the measurement system.
A picture of the real system at Kyoto University is shown in Fig.~\ref{fig:setup-pict}.
The light from a blue LED (Nichia NSPB500S) is diffused with a set of plastic plates and distributed to WLS fibers via an optical fanout.
We use Kuraray Y11(200)MS WLS fiber with 1~mm diameter, the same type as we use in the real T2K detectors.
The light intensity at each WLS fiber position is measured and corrected for when we calculate the photon detection efficiency.
The other end of the WLS fiber is connected to an MPPC using the optical connector developed for T2K detectors~\cite{Gomi}.

The signal from the MPPC is read out by the Trip-t ASIC~\cite{Trip-t} developed at Fermilab. 
Using two Trip-t chips, we are able to simultaneously measure 64 MPPCs.
The charge is recorded with VME-based electronics developed by our group~\cite{readout}.
The digital signal to control Trip-t and LED pulsing is generated using National Instruments PXI-6551.
The bias voltage, pulsing of the LED, VME readout and Trip-t control are synchronized and controlled by a Linux-PC.

The light source and MPPCs are kept inside a temperature controlled box.
The temperature inside the box is automatically controlled and changed to 25, 20, and 15$^\circ$C every 30 minutes after starting the measurement sequence.
Thus, data at three temperatures are automatically taken in 1.5~hours.

\section{Measurement procedure and performance of T2K-MPPC}
We record the charge from MPPC under two conditions: with and without light source.
The gate widths for integration are 200~ns and 800~ns for measurements with and without light, respectively.
Measurements are repeated with changing applied voltage (with 0.1~V step) and temperature (25, 20 and 15$^\circ$).
Based on the ADC distributions, we derive the basic performance of MPPC with the following method.
The measurement results with 5820 T2K-MPPCs are summarized in \cite{NDIP08-talk}.

\subsection{Gain and breakdown voltage}
\begin{figure}[tbp]
\begin{center}
\includegraphics[width=0.4\textwidth]{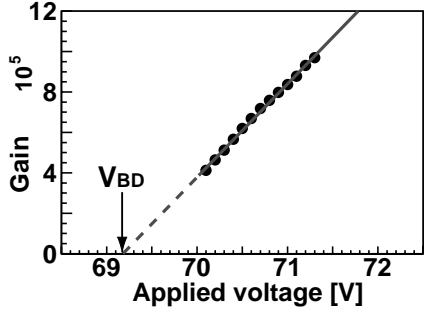}%
\caption{Determination of V$_\mathrm{BD}$ from gain-voltage extrapolation.}
\label{fig:Vbd}
\end{center}
\end{figure}%

The gain of MPPC can be easily measured because the pedestal, 1 photoelectron (p.e.), 2~p.e., ... peaks are well separated thanks to the high gain.
Measuring the charge corresponding to 1~p.e.\ and dividing it by the electron charge, we obtain the gain.
The gain $M$ depends linearly on the applied voltage $V$ as
$M = C (V-V_\mathrm{BD} ),$
where $C$ is the capacitance of each APD pixel and $V_\mathrm{BD}$ is the breakdown voltage above which the APD operates in the Geiger mode.
When we measure the gain as a function of the applied voltage, $V_\mathrm{BD}$ can be derived by linearly extrapolating the gain-voltage relation to the point where gain becomes zero (Fig.~\ref{fig:Vbd}).
The gain is about $5\times10^5$ at $\Delta V \equiv V-V_\mathrm{BD}=1.0$~V.

\subsection{Dark noise rate}
Dark noise rate can be measured from the data without external light,
by counting number of `hit' events and dividing it by the measurement time.
Because the MPPC has several effects that fake the photoelectron signal such as cross-talk and afterpulse, we always need to carefully evaluate the number of detected photoelectrons.
This is done as follows.
Assuming the number of \textit{true} detected photoelectrons, in absence of cross-talk and afterpulse, follows Poisson statistics, its mean $n$ can be estimated from the fraction of pedestal events among total triggered events $P(0)$ as \[n=-ln(P(0)).\]
Dividing $n$ by the gate width, dark noise rate is measured.
The typical dark noise rate for T2K-MPPC is about 0.8--1~MHz at $\Delta V=1.5$~V and 25$^\circ$C.

\subsection{Photon detection efficiency}
The photon detection efficiency (PDE) is relatively measured using a PMT (Hamamatsu R1818) as a reference.
The number of photoelectrons detected with an MPPC is derived from the fraction of pedestal events in the same way as explained above, with existence of weak external light.
The number of photoelectrons with PMT is measured for each WLS fiber to correct for the effect of non-uniform light distribution.
The light intensity of LED is monitored with a reference MPPC.
The PDE is defined as the number of photoelectrons detected by MPPC divided by that of PMT.

\begin{figure}[tbp]
\begin{center}
\includegraphics[width=0.4\textwidth]{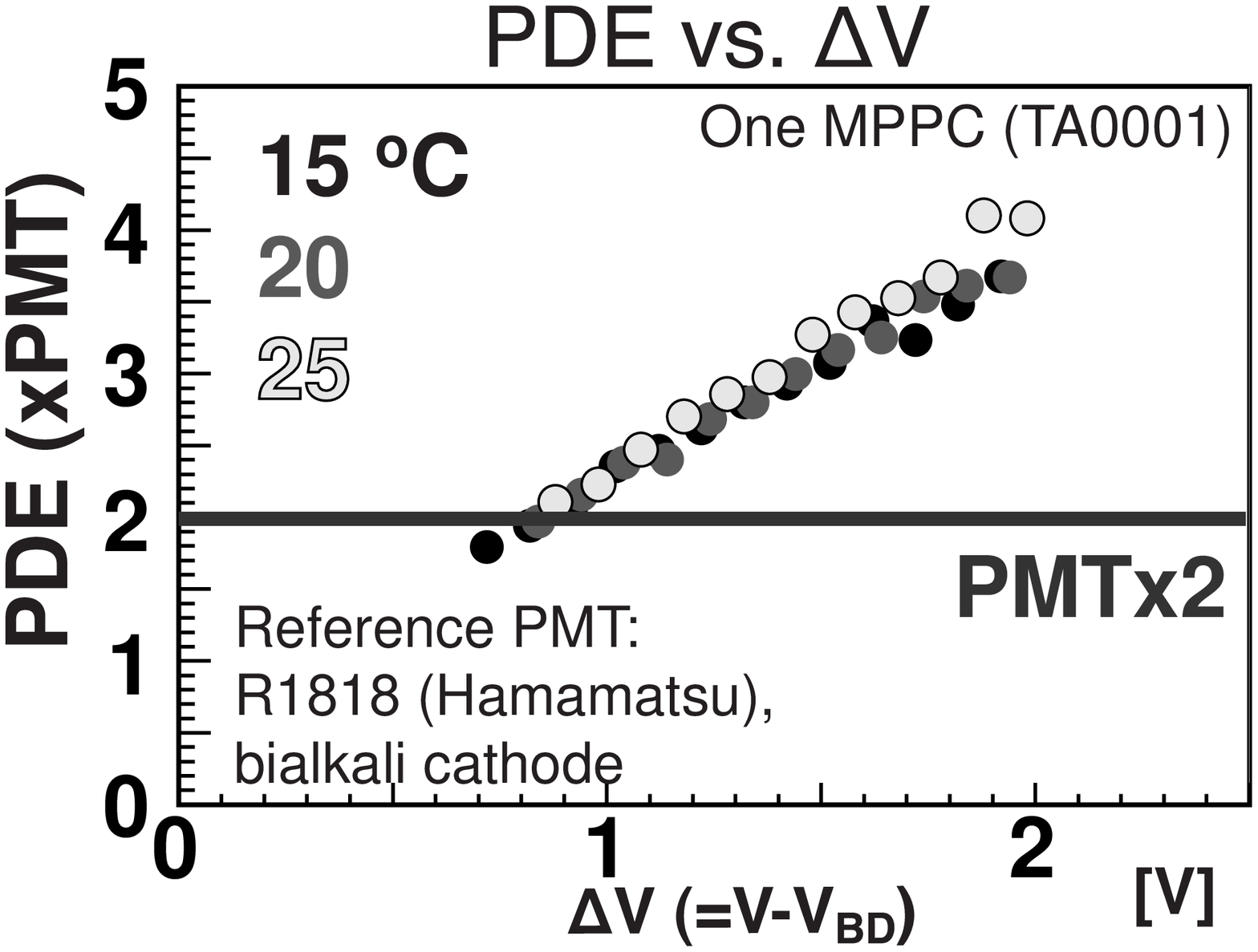}%
\caption{PDE.}
\label{fig:PDE}
\end{center}
\end{figure}%

Figure~\ref{fig:PDE} shows the measured PDE for an MPPC as a function of $\Delta V$ at 15, 20 and 25$^\circ$C.
It is seen that the temperature dependence is small, apart from the change of $V_\mathrm{BD}$ ($\d V_\mathrm{BD}/\d T \sim 50$mV/K~\cite{tempdep}).
The PDE of T2K-MPPC is much more than that of a usual PMT, satisfying our requirement.

\subsection{Cross-talk and afterpulse}
It is known that in an MPPC there are optical cross-talk to the neighboring APD pixel and afterpulse, both of which give an additional charge to the original photoelectron signal.
Because these additional charge also originates from the Geiger avalanche, it is indistinguishable from the original signal and makes e.g.\ 1~p.e.\ events to look 2~p.e.\ or more.
We are able to estimate the true number of 1~p.e.\ events, in absence of cross-talk and afterpulse, from the fraction of pedestal events (hence the mean value $n$) and Poisson statistics.
Comparing this number with the observed number of events at 1~p.e.\ peak and assuming the difference comes from the effects of cross-talk and afterpulse, we can estimate the rate of cross-talk and afterpulse from the ADC distribution.

\begin{figure}[tbp]
\begin{center}
\includegraphics[width=0.4\textwidth]{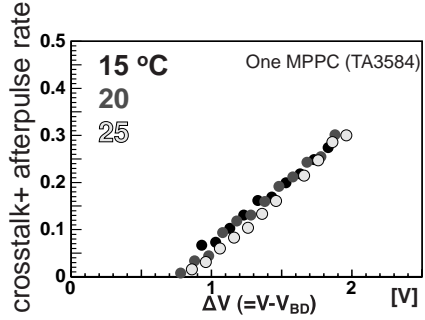}%
\caption{Cross-talk and afterpulse rate.}
\label{fig:CTAP}
\end{center}
\end{figure}%

Figure~\ref{fig:CTAP} shows the cross-talk and afterpulse rate of an MPPC, as a function of $\Delta V$ at 15, 20 and 25$^\circ$C.
Again, the temperature dependence is small if we correct for the change in $V_\mathrm{BD}$.
The cross-talk and afterpulse rate is about 0.05 (0.2) at $\Delta V=1.0$ (1.5)~V, which is acceptable for T2K but may be desired to improve for other applications.

\section{Conclusion}
We have developed an MPPC test system that can simultaneously characterize 64 MPPCs.
The method to evaluate the basic performance of MPPC such as gain, breakdown voltage, dark noise rate, photon detection efficiency, cross-talk and afterpulse rate, has been established.
With our test system, these parameters as functions of temperature and bias voltage are automatically measured.
We have measured the performance of T2K-MPPC and found it to satisfy our requirements.

\section*{Acknowledgments}
The authors are grateful to the solid state division of Hamamatsu Photonics for providing us test samples during the development.
The development of MPPC and its test system are supported by KEK Detector Technology Project, photon-sensor group.

\end{document}